\documentclass[10pt,conference]{IEEEtran}
\IEEEoverridecommandlockouts

\usepackage{cite}
\usepackage[square, numbers]{natbib}
\usepackage{amsmath,amssymb,amsfonts}
\usepackage{algorithmic}
\usepackage{graphicx}
\usepackage{textcomp}
\usepackage{xcolor}
\usepackage{multirow}
\usepackage{url}
\usepackage{booktabs}
\usepackage{tabularx}
\usepackage{nth}

\def\BibTeX{{\rm B\kern-.05em{\sc i\kern-.025em b}\kern-.08em
    T\kern-.1667em\lower.7ex\hbox{E}\kern-.125emX}}

\makeatletter
\newcommand{\linebreakand}{%
  \end{@IEEEauthorhalign}
  \hfill\mbox{}\par
  \mbox{}\hfill\begin{@IEEEauthorhalign}
}
\makeatother

\begin{document}

\title{Towards Detecting Cascades of\\Biased Medical Claims on Twitter\\
}
\author{
\IEEEauthorblockA{\begin{tabular}[t]{c@{\extracolsep{.7em}}c@{\extracolsep{.5em}}c@{\extracolsep{.5em}}c} 
Libby Tiderman  & Juan Sanchez Mercedes &  Fiona Romanoschi & Fabricio Murai\\
\textit{Dept.\ of Mathematics} & \textit{Dept.\ of Information Systems \&} & \textit{School of Information} & \textit{Data Science \& Computer Science} \\ 
\textit{University of Arizona} & \textit{Analytics, Bryant University} & \textit{Univerisity of Texas at Austin} & \textit{Worcester Polytechnic Institute} \\ 
Tucson AZ, USA & Smithfield RI, USA & Austin TX, USA & Worcester MA, USA\\
libbytiderman@arizona.edu & jsanchezmercedes@bryant.edu & fionaroman@utexas.edu & fmurai@wpi.edu
\end{tabular}}}

\maketitle

\begin{abstract}
\normalsize
Social media may disseminate medical claims that highlight misleading correlations between social identifiers and diseases due to not accounting for structural determinants of health. Our research aims to identify biased medical claims on Twitter and measure their spread. We propose a machine learning framework that uses two models in tandem: RoBERTa to detect medical claims and DistilBERT to classify bias. After identifying original biased medical claims, we conducted a retweet cascade analysis, computing their individual reach and rate of spread. Tweets containing biased claims were found to circulate faster and further than unbiased claims.
\end{abstract}


 
\section{Introduction}
Social media platforms serve as effective tools for spreading information on public health and social issues. They function as outlets for people to share opinions, experiences, and converse with others they may never communicate with otherwise. Platforms are also susceptible to misuse: individuals can spread hateful, biased, and erroneous information, particularly concerning medical matters \cite{article}. Anyone can post medical claims on social networks, making it difficult for users to distinguish opinions, objective facts, incorrect or biased information \cite{Trethewey2019}. Widespread biased medical information can be particularly harmful to people of certain racial groups, gender identities, sexualities, etc~\cite{Dori-Hacohen2021}. The potential influence of such medical claims on users may result in people being discouraged from seeking professional healthcare, and can even reinforce biases in how doctors perceive and treat patients.

Claims are the central part of an argument, which gives valuable insight into the author's views and is likely to stand out to other Twitter users \cite{Stab2017}. Discerning the argument a user is making on Twitter can provide insight into public perception and trends. The ability to ascertain the presence of gender bias in medical claims on social media is crucial for identifying communities that may be disproportionately affected by that content. As an example, a tweet stating ``\textit{98\% of autism in men is just depression}'' can invalidate the experiences of men diagnosed with autism or others seeking information about depression. In both instances, this could dissuade an individual from seeking help or influence general opinion of these conditions. Another essential component of determining the impact of biased medical claims is measuring how much and how fast these tweets spread online. A user is able to repost another’s original tweet through the retweet function, displaying it on their followers’ timeline and own profile. This flow of content corresponds to an information cascade on Twitter. Through constructing cascades of retweets, we are able to assess the visibility, as retweet count has been shown to be positively correlated with follower count \cite{sakamoto2015analysis}. 

Claim detection on social media has been extensively studied, including in the context of biomedical facts \cite{Sundriyal2022, Gupta2021, wuhrl2021}. Identifying biased information in medical literature has been addressed 
\cite{Markowitz2022} and several studies have incorporated bias detection modules into hate speech detection models \cite{NASCIMENTO2022117032, Mozafari2020}. The spread of misinformation on social media has been investigated \cite{Vosoughi2018, WASZAK2018}, however there are no studies that measure the diffusion of biased medical information in online networks. As a first step in that direction, we quantify the dissemination of gender-biased mental health claims in order to gain insights into the reach of harmful information on Twitter. Instead of traditional user-centric metrics often used to evaluate cascade patterns, our study examines the relation of a medical tweet’s bias to its spread. In this study, we present a machine learning framework for identifying biased medical claims on Twitter that consists of: detecting medical claims in tweets containing both a gender and mental health keyword utilizing RoBERTa, assessing the level of bias in medical tweets using DistilBERT, and quantifying the dissemination of those selected tweets. 

We trained a RoBERTa model for medical claim detection using labeled Twitter data from \citet{Gupta2021} and fine-tuned it to fit the needs of this task. A subset of 300 tweets was manually annotated to specify if a medical claim was present in each tweet to serve as a validation set for the claim detection models. DistilBERT was utilized to detect bias in medical claims by training it on medical curriculum annotated for various biases. Cascades were constructed of the top 10\% of tweets most likely to be biased and bottom 10\% based on the predicted probabilities calculated by the bias detection model. Finally, we compared the cascades and found that biased claims spread further and faster than unbiased claims.

\medskip

\section{Related Work}
We review the related literature regarding claim detection, hate speech detection, and cascade analysis. A variety of methods were employed from these works and our training dataset for the medical claim detection model came from the \citet{Gupta2021} in particular.

\subsection{Claim Detection}

Researchers have explored various approaches for claim detection, including isolating entire claims rather than simply classifying text as containing claims or not. \citet{Sundriyal2022} presented a model named DaBERTa specifically tailored for claim isolation. Their work demonstrated that DaBERTa surpassed the performance of both BERT and RoBERTa in this task. However, it is essential to note that the focus of our study did not involve claim isolation, and therefore, we opted to work with RoBERTa, which offered easier implementation for our particular needs. 

\citet{wuhrl2021} took a different route by targeting claim detection in the domain of biomedical tweets. They manually annotated 1200 biomedical tweets as either claims or non-claims. The annotated dataset was then utilized to evaluate the performance of four distinct models: Gaussian Naive Bayes, logistic regression, BiLSTM, and BERT. Surprisingly, the results indicated that the simpler models (Gaussian Naive Bayes and logistic regression) outperformed the more complex models. As discussed by the authors, this results is likely due to the small dataset size (800 Tweets), which limits the potential of the more sophisticated models. 

In another related study, \citet{Gupta2021} conducted claim detection in the context of social media platforms, specifically focusing on Twitter. They annotated 9,894 tweets, classifying them as either claims or non-claims, and subsequently trained their innovative model, LESA, for this task. LESA exhibited slightly improved performance compared to BERT.  We found it difficult to reproduce LESA’s results, therefore, we proceeded with fine-tuning a RoBERT model instead, for achieving our research objectives.

\subsection{Hate Speech Detection}

Many researchers have been studying hate speech detection models, however few focus on bias specifically. \citet{hatespeech2023ht} trained HateBERT with six commonly employed datasets for hate speech detection tasks. Two of these had a high frequency of gendered terms, however the model didn’t generalize well between them. Although HateBERT performed well for hate speech detection, we decided not to utilize it for bias classification because it was trained on reddit data, which is very different from the training data consisting of medical curriculum that was available to us. 

Other studies have begun to implement bias mitigation modules into their hate speech detection models. \citet{Mozafari2020} attempted to mitigate racial bias in their training data by reweighting biased text before feeding it through a pre-trained BERT model. While this is a promising approach, this is not applicable to our task of bias classification. Similarly, \citet{NASCIMENTO2022117032} implemented a bias mitigation module that replaced gendered words with the key <identity> so that the bias detection model would predict based on other inflammatory words in the sentence. Since we collected tweets that specifically contained gender keywords, it would not make sense for us to classify bias based on the presence of a keyword or to replace them with a key.

\medskip

\section{Overview of Proposed Framework}

We propose a new framework for identifying medical claims, quantifying the associated bias, and evaluating the resulting cascades as shown in Figure 1. The initial stage involves determining the tweets within our dataset that contain medical claims using a claim detection model. After filtering out all of the non-claims, we feed the remaining tweets into a bias detection model to assign them bias probability. Last, we identify all the retweets of the original medical claims in our dataset and build their respective cascades.

\begin{figure}
    \centering
    \includegraphics[width=.9\linewidth]{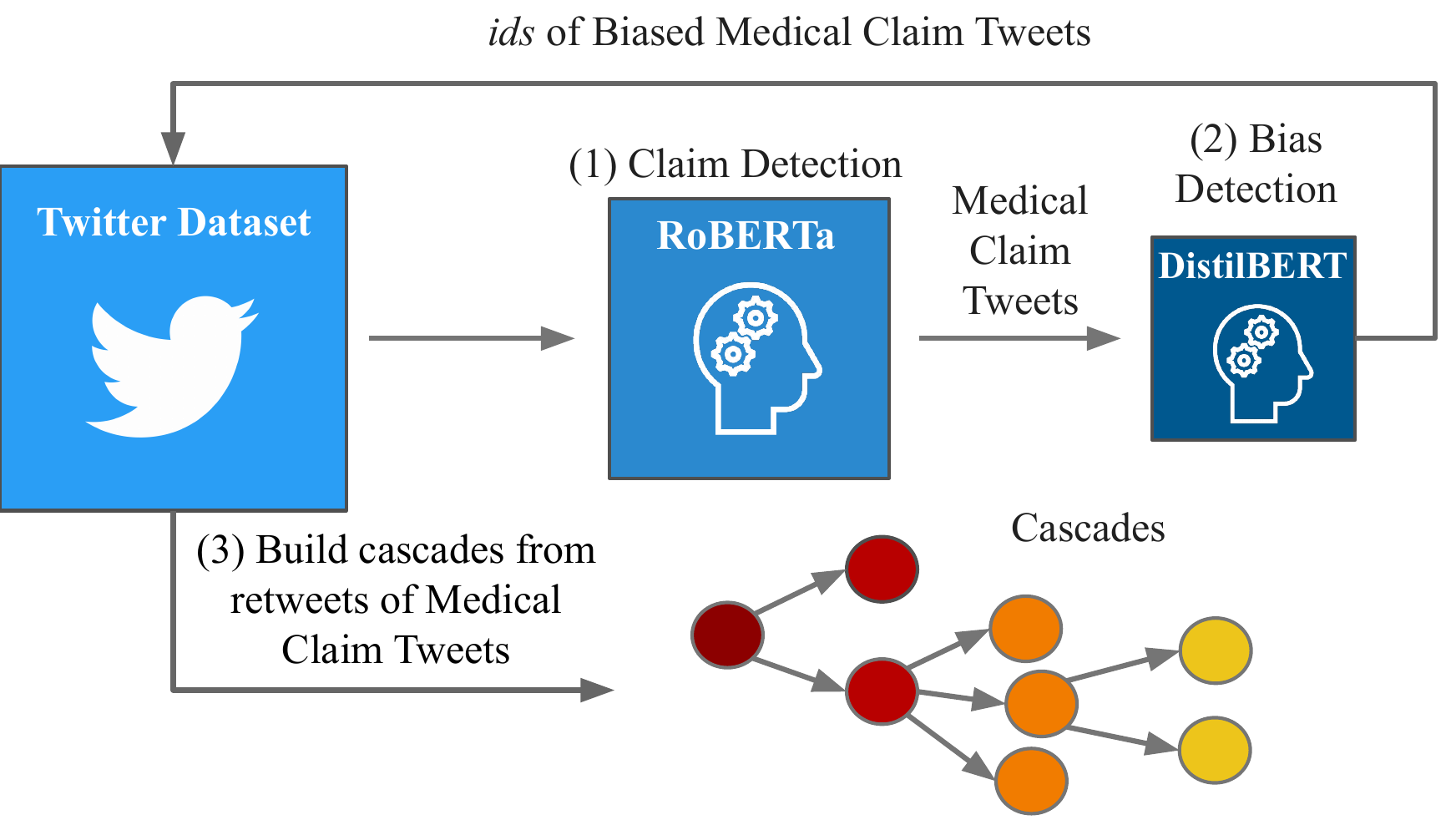}
    \caption{Proposed Framework consists of three steps: (i) identifying original tweets that contain medical claims, (ii) computing a bias score, and (iii) building cascades from the dataset.}
    \label{fig:enter-label}
\end{figure}

\medskip

\section{Methodology}

After collecting more than 6 million tweets containing both gender and mental health keywords, we trained a RoBERTa-base model on annotated Twitter data from \citet{Gupta2021} to identify claims. Subsequently, we annotated 300 tweets from our dataset to distinguish between those containing a medical claim and those without one to serve as a validation set for the RoBERTa model. After classifying the remaining tweets in our dataset using RoBERTa, we trained a DistilBERT model on formal medical curriculum annotated for various biases, assigning a bias probability score to each original tweet containing a medical claim. Finally, we constructed cascades of the top 10\% most likely to be biased to the bottom 10\%.

\subsection{Twitter Dataset}
A large dataset was collected comprising more than six million tweets, which was collected through a series of Twitter API queries containing both a mental health and a gender related keyword. The dataset comprised 57.5\% retweets, 26\% replies, 3\% mentions/quotes and 13.5\% original tweets as shown in Figure 2. The average number of likes of a tweet was 4.85 and the average number of views was 193.92. The oldest tweet in our dataset was from December 2007 and the newest tweet was from April 2023. The vast majority of tweets were not geotagged (0.6\%), although the top places of tweet origin were Los Angeles, Florida, Houston, Manhattan and Chicago in that order. To ensure that the dataset was suitable for further analysis, we developed a preprocessing function that cleaned the text and removed hashtags, URLs, and stopwords.

\begin{figure}
    \centering
    \includegraphics[height=1.5in]{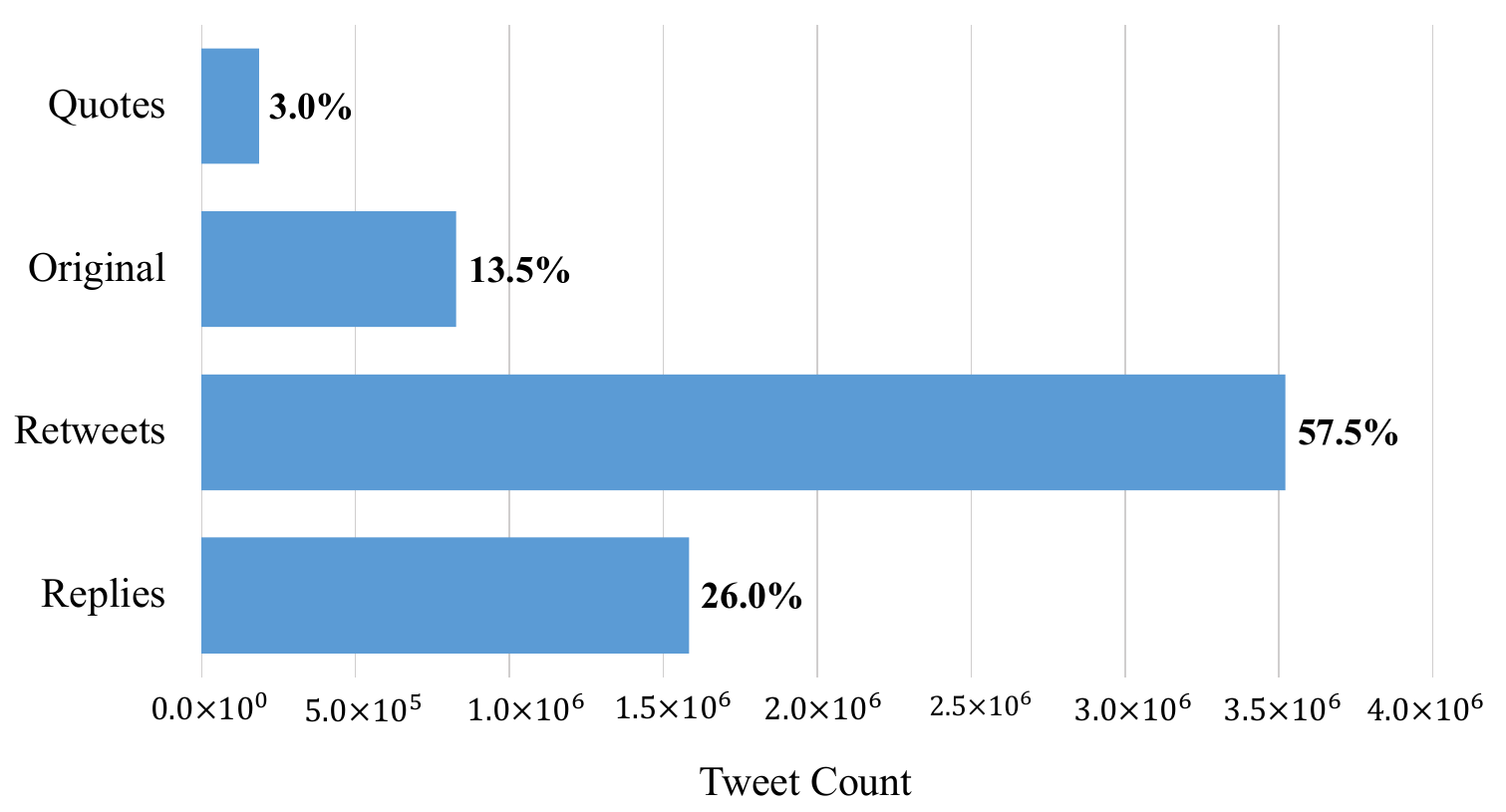}
    \caption{Breakdown of interaction types in our Twitter dataset.}
    \label{fig:enter-label}
\end{figure}

\subsection{Medical Claim Detection Model (CDM)}
Next, we trained both a BERT-uncased and a RoBERTa-base model (transformer-based language model) for claim detection using annotated data from \citet{Gupta2021}. Due to the heavy class imbalance seen in Table~\ref{tab:LESA_stats}, we adjusted the model weights to 6.93-1.00 (weight = positive/negative instances = $6977/1007 = 6.93$). However, when attempting to train the entire model, we observed that it overfitted very quickly. To avoid this, we froze the pre-trained layers and trained only the last two classification layers for 20 epochs. Once the learning rate converged, we trained the whole model for two additional epochs, using a constant learning rate of $10^{-5}$.

\begin{table}
\caption{Summary of Training Data for Claim Detection Model}
\label{tab:LESA_stats}
\begin{center}
\begin{tabular}{ c|rr|r }
\toprule
  & Claim & No Claim & Total \\ 
 \midrule
 Train & 6,877 & 1,007 & 7,984 \\  
 Test & 1,745 & 252 & 1,897 \\   
\bottomrule
\end{tabular}
\end{center}
\end{table}

\subsection{Labeling an Evaluation Set of Tweets}
To assess the claim detection models' performance on our dataset, we annotated 300 randomly sampled tweets as medical claims for validation. We initially annotated 100 tweets, then had two medical doctors provide their own annotations. After reviewing their annotations, we discussed and resolved any disagreements, then proceeded with labeling the remaining 200 tweets. Any disputes of these annotations were settled by a single doctor, which resulted in 300 tweets for validation. 

\subsection{Calibrating the CDM}
We performed a threshold selection to assess the performance of the RoBERTa and BERT model's predictions at different threshold values for our validation dataset. Since our goal was to label as many true positive cases (claims) as possible, we decided that optimizing the precision (proportion of observations predicted as positive that are, in fact, positive) for the claims would be the best measure for evaluating our model. This leads to a more conservative model that labels only the tweets with a high probability of being a claim as a positive case. By analyzing the ROC curve for the evaluation data, we chose a threshold value that yielded the highest precision before the recall (proportion of positive observations predicted as such) started to drop significantly. We chose the threshold ($\tau = 0.91$) so that only tweets with a predicted probability above that threshold were labeled as a claim. After selecting a new threshold for both models, we found that the RoBERTa outperformed the BERT-uncased model. The RoBERTa had a precision of 0.80 for the positive cases while the BERT only had 0.33, as seen in Table~\ref{tab:validation}. Given this, we decided to move forward with the RoBERTa for labeling the rest of our dataset. 

\begin{table}[t]
\caption{RoBERTa and BERT Comparison on Validation Set}
\label{tab:validation}
\begin{center}
\begin{tabular}{cc|ccc}
\toprule
\multicolumn{2}{c|}{} & {Precision}& {Recall}& {F-1 Score} \\
\midrule
\multirow{ 2}{*}{RoBERTa} & Non-Claim & 0.91&  1.00& 0.95\\
 &Claim& 0.80& 0.12&0.22\\
\midrule
\multirow{ 2}{*}{BERT}&Non-Claim&0.90&0.99&0.94\\
&Claim&0.33&0.06&0.11\\
\bottomrule
\end{tabular}
\label{tab1}
\end{center}
\end{table}

\subsection{Bias Classification Model (BDM)}

To identify biases present in the tweets related to medical claims, we created a Bias Classification Model. We use a corpus of 224 text excerpts taken from medical school course materials and annotated by medical students as part of the Bias Reduction in Curricula Content (BRICC) project run by the University of Washington. Excerpts were initially selected for containing potentially biased claims and then labeled as: non-biased or different bias categories (e.g., age, gender, race). All biased instances (83) are mapped to positive labels (i.e., label 1). We augment this corpus with an equal number of hard negative instances by finding, in the course materials, the most similar sentence to each of the positive examples. Our final dataset contains 83 positive and 365 negative instances. We use a stratified 80-10-10\% split based on the fine-grained bias categories to ensure that all types of biases are present in each split. 
Among transformer based models, we opted for DistilBERT instead of RoBERTa because our dataset consists of medical curriculum as opposed to social media content. DistilBERT was applied to the tweets that contained medical claims to estimate the probability that they were biased. 

\subsection{Cascade Construction}

A cascade refers to the hierarchy of messages posted on Twitter associated with an original tweet, which can be either replies, retweets, or quotes. Due to the fact that data collection was performed by querying specific keyword pairs, replies and quotes that do not contain any of the keyword pairs are not present in the collected data. For this reason, we only use retweets to build the cascades. From the large Twitter dataset, an edgelist was constructed connecting the tweet id of the prospective tweet and the tweet id of its parental node. We chose to analyze cascades that started with a completely original medical claim (is not a mention, retweet, or reply). Then, we conducted a breadth-first search in NetworkX~\cite{networkx} of those nodes to construct their individual cascades. In our analysis, we compare the top 10\% (most likely biased) to the bottom 10\% (least likely biased) as predicted by BDM, henceforth referred to as biased and unbiased cascades.

\medskip

\begin{table}
\caption{Examples of medical claims and corresponding CDM and BDM outputs. \nth{1} is in 10\% \textcolor{red}{most biased}, \nth{2} is in 10\% \textcolor{blue}{least biased}.}
\label{tab:examples}
\begin{tabularx}{\columnwidth}{Xcc}
\toprule
\multicolumn{1}{c}{Example Tweet} & P(claim) & P(bias) \\
\midrule
\textit{There are only 2 genders. Gender dysphoria is a mental health issue caused by environmental estrogen and indoctrination...} & 0.91 & \color{red}{0.68} \\
\hline
\textit{elevated paternal glucocorticoid exposure alters the small noncoding RNA profile in sperm and modifies anxiety and depressive phenotypes in the offspring} & 0.94 & \color{blue}{0.32}\\
\bottomrule
\end{tabularx}

\end{table}

\begin{figure*}
\begin{minipage}[t]{0.32\textwidth}
  \includegraphics[width=\linewidth]{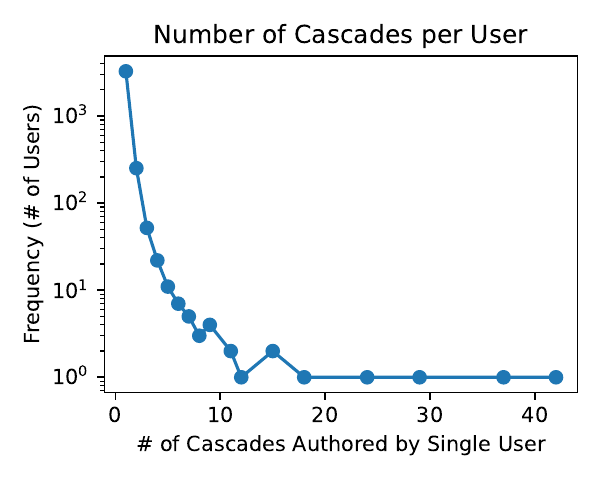}
  \caption{Cascade Authorship (all cascades)}
  \label{fig:third}
\end{minipage}%
\hfill
\begin{minipage}[t]{0.31\textwidth}
  \includegraphics[width=\linewidth]{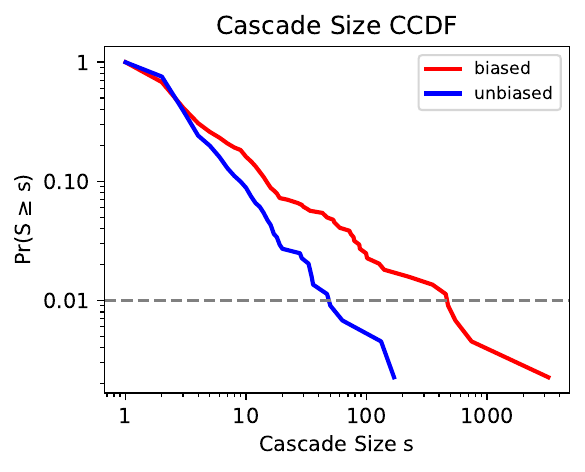}
  \caption{Cascade Size Comparison}
  \label{fig:first}
\end{minipage}%
\hfill 
\begin{minipage}[t]{0.31\textwidth}
  \includegraphics[width=\linewidth]{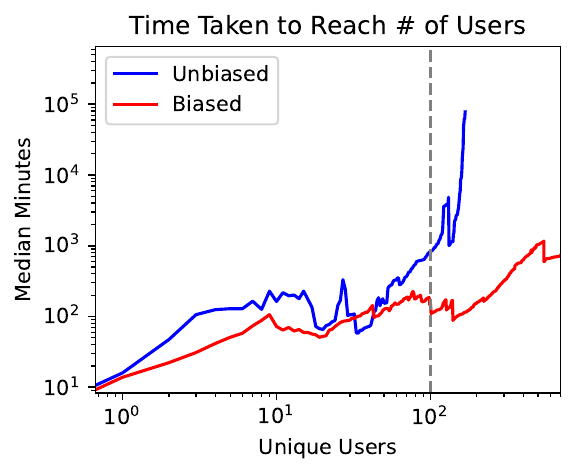}
  \caption{Cascade Velocity Comparison}
  \label{fig:second}
\end{minipage}%
\end{figure*}

\section{Results}
In this section, we present the results of the Claim and Bias Detection Models, followed by the characterization of the biased and unbiased cascades.
\subsection{Claim and Bias Detection Models (Example)}
To illustrate the outputs of each model, we show in Table~\ref{tab:examples} two examples of tweets from our dataset and the probabilities that each contained a medical claim and was biased. Based on our threshold value of $\tau = 0.91$, both tweets were labeled by RoBERTa as medical claims. The first claim had a 68\% chance of being biased according to our bias detection model and is in the top 10\% of original tweets most likely to be biased. As we can see, this tweet makes a strong exclamation that there are only two genders, then implies that the mental health condition, gender dysphoria, is the cause of people identifying outside of a binary gender. The statement is incorrect and biased, as identifying as a different gender does not equate to having gender dysphoria~\cite{Richards2016}. Conversely, the second claim has a much lower probability of being biased at 32\% and is located in the bottom 10\%. In fact, the sentence does not exhibit a clear gender bias.


\subsection{Cascade Analysis}
We characterized the cascades based on:  (i) number of cascades initiated per user, (ii) size (number of unique users), (iii) velocity, measured as the median time in minutes to be retweeted a certain number of times.

Figure~\ref{fig:third} shows that the distribution of the number of cascades per user is heavy-tailed: some users authored much more medical claims than the average.
Within our dataset, 14.71\% of users authored two or more cascades.

The average size or number of nodes in a cascade was 17.77 tweets. Figure~\ref{fig:first} indicates that, on average, biased cascades were larger, making them more likely to spread to a larger number of users. In particular, the top 1\% of the biased cascades reach at least 500 users, while the top 1\% of the unbiased ones reach 40 or more users. 

 Last, Figure~\ref{fig:second} shows that, in general, biased cascades take much less time than the unbiased ones to be retweeted the same number of times. For instance, to be retweeted 100 times, the median biased cascade takes 145.31 minutes, whereas the unbiased counterpart takes 822.43 minutes. 

\medskip

\section{Conclusions and Future Work}

We made the first step towards studying  the propagation of biased medical claims on Twitter using a combination of claim detection, bias classification and network analysis techniques. The use of two separate classifiers circumvents the need to have a large dataset labeled for biased medical claims, allowing us to leverage two existing datasets: the first containing tweets labeled as medical claim or not a medical claim, and the second containing excerpts from medical curriculum labeled for the presence of bias. Based on our experiments, we concluded that using RoBERTa as a base for training the CDM (claim detection model) and DistilBERT as a base for the BDM (bias detection model) yields satisfactory results. We use the CDM and BDM in tandem to identify original tweets that contain biased medical claims. In our study, we consider gender bias in mental health-related claims. We observed that biased cascades exhibited greater reach and faster dissemination than unbiased medical claims. In the future, we plan to extend this study to other types of biases and diseases, using the proposed framework.

RoBERTa suited our scope particularly well, as it is pre-trained on a corpus that includes Reddit content, which explains why it performs better on our social media data \cite{liu2019roberta}. DistilBERT's training on formal medical curriculum made it suitable for bias detection. However, one limitation of our study is that we were not able to validate BDM on social media data. Also, since we were unable to collect user data, we couldn't perform a more in-depth social network analysis as many cascade metrics are correlated with user centrality in the network \cite{Suh2010}.
Reinstating academia-related API partnerships is vital for continued research in this field.

One promising direction for future work is the design of specific ML models or training procedures for bias detection based on small samples. Additionally, applying the proposed framework to other social media platforms would provide valuable insights into how certain viewpoints and communities interact across networks.

The outcomes of our study have significant implications for addressing misinformation and promoting the dissemination of accurate medical information within online social networks. The findings contribute to the development of effective strategies and interventions to mitigate the adverse effects of biased medical information. Ultimately, different medical issues are prevalent in certain communities. Tracking these issues and their corresponding biases could help society, social network leaders and government officials grasp the common perception and counteract these issues.

\section*{Acknowledgments}

This material is based upon work supported in part by the National Science Foundation REU Site Grant 1852498 and Grant IIS-2147305. We thank Shannon Song for her mentorship, and Chiman Salavati and Shiri Dori-Hacohen for their contributions. Any opinions, findings and conclusions or recommendations expressed in this material are those of the author(s) and do not necessarily reflect the views of the National
Science Foundation.


\bibliography{citations}

\end{document}